\documentclass{IEEE_lsens}
\usepackage{textcomp}

\RequirePackage[colorlinks=true, allcolors=blue]{hyperref}

\usepackage[noadjust]{cite}

\ifCLASSINFOpdf
  \usepackage[pdftex]{graphicx}
  \DeclareGraphicsExtensions{.pdf,.jpeg,.png}
\else
\fi
\usepackage[T1]{fontenc} 
\usepackage{amsmath}
\usepackage[cmintegrals]{newtxmath}
\usepackage{bm}

\usepackage[capitalize,noabbrev]{cleveref}
\usepackage{multirow}
\usepackage[final]{changes} 

\usepackage{array}
\usepackage{url}
\ifCLASSINFOpdf
\else
\fi
\providecommand{\hypersetup}[1]{\relax}

\hypersetup{pdftitle={Pixel Watch: Robust Heart Rate Sensing from Multipath PPG and On-Device Deep Learning Trained on 10,000 hours of Free-Living and Fitness Data},
pdfsubject={Sensor Applications},
pdfauthor={D. Roggen},
pdfkeywords={Wearable physiological sensing, heart rate sensing, PPG, on-device deep learning, validation study, Google Pixel watch}}
\begin{document}

\markboth{Vol.~x, No.~x, May~20xx}{0000000}

\IEEELSENSarticlesubject{Sensor Applications}

\title{Pixel Watch: Robust Heart Rate Sensing from Multipath PPG and On-Device Deep Learning Trained on 10,000 hours of Free-Living and Fitness Data}



%
\author{\IEEEauthorblockN{Daniel Roggen\IEEEauthorrefmark{1,2}\IEEEauthorieeemembermark{1}, 
Megan Walker\IEEEauthorrefmark{1},
Yojan Patel\IEEEauthorrefmark{1}, 
Shyam Tailor\IEEEauthorrefmark{1}, 
Dimitris Spathis\IEEEauthorrefmark{1,3},
Matt Wimmer\IEEEauthorrefmark{1}, 
Brennan Garrett\IEEEauthorrefmark{1},
Dan Howe\IEEEauthorrefmark{1},
Abhinuv Pitale\IEEEauthorrefmark{1}, 
Hamed Vavadi\IEEEauthorrefmark{1}, 
Tien Le\IEEEauthorrefmark{1}, 
Steve Diamond\IEEEauthorrefmark{1}, 
Oleksiy Vyalov\IEEEauthorrefmark{1}, 
Vik Sharma\IEEEauthorrefmark{1}, 
Pete Richards\IEEEauthorrefmark{1}, 
Tracy Giest\IEEEauthorrefmark{1}, 
Erika Siegel\IEEEauthorrefmark{1}, 
Tuan Phan\IEEEauthorrefmark{1},
Sam Mravca\IEEEauthorrefmark{1},
Derrick Vickers\IEEEauthorrefmark{1}, 
Benjamin Stone\IEEEauthorrefmark{4}, 
Katarina Vukosavljevi\'{c}\IEEEauthorrefmark{1},
Justin Phillips\IEEEauthorrefmark{1},
YongSuk Cho\IEEEauthorrefmark{1},
Stefanie Hollidge\IEEEauthorrefmark{5}, 
Antony Siahaan\IEEEauthorrefmark{5},
Soren Brage\IEEEauthorrefmark{5},
Shwetak Patel\IEEEauthorrefmark{1,6}, 
Robert Harle\IEEEauthorrefmark{1,3}}
\IEEEauthorblockA{
\IEEEauthorrefmark{1}Google Research, Mountain View, CA, 94043, USA\\
\IEEEauthorrefmark{2}School of Engineering \& Informatics, University of Sussex, Falmer, UK\\
\IEEEauthorrefmark{3}Department of Computer Science and Technology, University of Cambridge, Cambridge, UK\\
\IEEEauthorrefmark{4}Akraya, Sunnyvale, CA, 94089, USA\\
\IEEEauthorrefmark{5}IMS Epidemiology, University of Cambridge School of Clinical Medicine, Cambridge, UK\\
\IEEEauthorrefmark{6}Department of Electrical \& Computer Engineering, University of Washington, Seattle, USA\\
\IEEEauthorieeemembermark{1}Member, IEEE}%
\thanks{Corresponding author: D. Roggen (e-mail: roggen@google.com).\protect}
\thanks{Associate Editor: xxxx xxxx.}%
\thanks{Digital Object Identifier 10.1109/LSENS.20xx.0000000}}
%
%
%

\IEEELSENSmanuscriptreceived{Manuscript received Month nn, 20xx;
revised Month nn, 20xx; accepted Month nn, 20xx.
Date of publication Month nn, 20xx; date of current version Month nn, 20xx.}

\IEEEtitleabstractindextext{%
\begin{abstract}[block6]

The Pixel Watch 2 (PW2) is the first Google smartwatch to combine multipath photoplethysmography (PPG) with deep learning-based heart rate inference, designed to significantly improve sensing accuracy during motion-heavy activities. 
The device processes 10 optical channels using an on-device, 15-layer temporally dilated convolutional neural network ($\sim$300K parameters) to yield a 1 Hz heart rate output. Crucial to this model's performance was its training on a massive dataset comprising 10,000 hours of data from 962 participants, curated from a broader corpus of controlled and free-living activities. 
We evaluated the PW2's sensing performance across two independent validation sets: an in-house fitness dataset (229 participants, 250 hours) and an external free-living dataset (27 participants, 1000+ hours). 
The system achieved 95\% Limits of Agreement of -10.34 to 8.66 BPM during exercise and -6.57 to 7.48 BPM during free-living activities, demonstrating substantially tighter error margins than previous Google devices. 
Finally, we discuss key design lessons, emphasizing that large-scale deep learning was instrumental in fully leveraging multipath PPG hardware over traditional signal processing approaches.

\end{abstract}

}



\maketitle

\section{Introduction}
\label{sec:introduction}




\IEEEPARstart{Wearables } are increasingly used in large-scale health studies, such as UK Biobank \cite{Biobank17} or ``All of Us'' \cite{AllOfUs19} as well as by individuals to guide their health decisions.
Therefore, it is important to understand how these devices were designed and evaluated to measure physiological parameters. 
%
Our contributions are:
\begin{itemize}
\item An overview of validation studies of consumer wearable heart rate sensing devices, highlighting the challenge of measurements with motion artifacts in \cref{sec:soa_validation}, and an overview of advances to improve robustness to motion artifacts through multipath and multiple wavelength and deep learning in \cref{sec:soa_advances}. 
\item A description of the PW2 PPG hardware and of the deep learning model used to continuously infer heart rate 
in \cref{sec:hwsw}. 
\item A description of the datasets in~\cref{sec:dataset}, emphasizing the unique scale and diversity of the training data, and detailing the unseen evaluation datasets including a fitness activity dataset to evaluate the effects of motion artifacts and a cohort dataset in free-living conditions to understand real-world performance. 
\item We report on performance using well established measures (MAE, MAPE, LoA, Bland-Altman plots) in \cref{sec:evaluation}.
\item We conclude contrasting the system performance to previous studies, highlighting the step-change in quality compared to prior Google devices, and highlight the lessons learned for future sensing system designs in \cref{sec:conclusion}. 
\item We conclude that the use of a deep model was instrumental to leverage multipath PPG compared to our prior work with classical signal processing. 
It also enabled us to systematically explore \replaced{multipath}{multipaths} options while reducing engineering effort, thus paving the way for future more complex optical arrangements.
\end{itemize}

\section{Related work}
\label{sec:soa}


\subsection{Validation of wristworn HR measurements}
\label{sec:soa_validation}

Continuous heart rate monitoring via wrist-worn PPG has become mainstream but ensuring accuracy during dynamic activities remains a challenge.
Indeed, one of the first validation studies of a consumer wearable evidenced this under conditions of moderate and vigorous physical activity (MVPA) in the Fitbit Charge HR \cite{Gorny17}.
Since then, many validation studies have been published, and a common limitation is often small sample size, with often fewer than 50 participants and only a few hours of data of prescribed activities.
We argue that a comprehensive evaluation of performance requires multiple settings including prescribed activities as well as real-life evaluation, which we follow in this work, including more participants and larger data volume (\cref{tab:hr_soa}).
%
\replaced{Our survey shows that an ECG chest strap is generally used as a criterion device (\cref{tab:hr_soa}), and we follow this practice in this work.}{Generally an ECG chest strap  is used as criterion device.}
%
Evaluation best practices typically suggest reporting mean absolute error (MAE), mean absolute percentage error (MAPE), and Bland-Altman (BA) plots with limits of agreement (LoA) \cite{Nelson20}.
%



\begin{table}[ht!]
\tiny
\centering
\caption{%
Compared to previous wearable heart rate validation studies, our work uses more participants and hours of data.
ICC = Intraclass correlation coefficient; CV=Coefficient of variation; Pearson=Pearson correlation; sTEE=standardised typical error of the estimate%
. }
\label{tab:hr_soa}
{
\setlength{\tabcolsep}{1pt}
\begin{tabular}{|p{0.04\linewidth}|p{0.19\linewidth}|p{0.11\linewidth}|p{0.4\linewidth}|p{0.20\linewidth}|}
\hline
Ref & Device & Criterion & Evaluation dataset & Evaluation\\
\hline
\cite{Gorny17} & Fitbit Charge HR & Polar\textsuperscript{\textregistered} H6 & n=10, 3-6 hours/person + 1 month free living & ICC, BA\\
\cite{Bai21} & Fitbit Charge 2\&Alta, Apple Watch 2 & Polar\textsuperscript{\textregistered} H7 & n=48, 24 hours: free-living & MPE, MAPE, RMSE, equivalence test, BA\\
\cite{HajjBoutros23} & Fitbit Sense, Apple Watch 6, Polar\textsuperscript{\textregistered} Vantage V & Polar\textsuperscript{\textregistered} H10 & n=60, 50mn/person: sitting, walking, running, resistance exercises, cycling & Pearson, CV, MAPE, BA, sTEE\\
\cite{Muggeridge21} & Polar\textsuperscript{\textregistered} OH1, Fitbit Charge 3 & Polar\textsuperscript{\textregistered} H10 & n=20, 15-minute sedentary, 10-minute cycling, treadmill test (18-42mn) & Bias, LoA, correlation coefficient, MAPE \\
\cite{Lima24} & Samsung Galaxy Watch 4 & Polar\textsuperscript{\textregistered} H10 & n=14, 40mn/person: lab treadmill and cycling & Dropout rate, performance index, bias, precision, MAE\\
\cite{Rho23} & Samsung Galaxy Watch 5 & Shimmer3 ECG &  n=20, 2mn/person, resting and walking  & Pearson\\
\cite{Nissen22} & Fitbit Charge 4, Samsung Galaxy Watch Active2 & Holter ECG &n=23, 15mn/person: resting, sedentary, low-intensity and high-intensity & Bias, MAE, BA\\
\cite{Nuuttila22} & Polar Vantage V2 &  Polar\textsuperscript{\textregistered} H10 & n=39 at rest 5mn/subject, n=29 during sleep with 4h/subject healthy, recreationally endurance-trained & MAE, MAPE, Pearson, ICC, Lin’s concordance correlation coefficients, BA \\
\cite{Montalvo23} & Apple Watch 6 and 7 & Polar\textsuperscript{\textregistered} H10, ECG & n=44: treadmill test incl. rest, walk, run & ICC, BA\\
\cite{Damasceno22} & Garmin Forerunner 735XT &  ErgoMET ECG & n=28, 30mn/person: rest, walk, run & MAE, MAPE, BA\\
\cite{Duking20} & Apple Watch 4, Polar Vantage V, Garmin Fenix 5, Fitbit Versa & Polar H7 & n=25, 33mn/person: sitting, walking, running & sTEE, CV, Pearson, bias \\
\hline
\multirow{2}{=}{This work} & \multirow{2}{=}{Pixel Watch 2} & Polar\textsuperscript{\textregistered} H10 & n=229, 65mn/person: fitness activities & \multirow{2}{=}{MAE, MAPE, BA, LoA}\\
 &  & Actiheart 4 & n=27, 48hrs/person: free-living & \\
\hline
\end{tabular}
}
\end{table}

\subsection{Enhancing robustness to motion artifacts}
\label{sec:soa_advances}

One of the first uses of deep learning to handle motion artifacts was presented in \cite{Reiss19}, yielding a model with 26K parameters operating on the PPG spectrogram and acceleration data.
Since then deep learning has become the approach of choice in the literature to infer heart rate from PPG, and many variations have been explored, such as using wrist-band pressure instead of acceleration to combat the motion artifacts \cite{Mehrgardt21}, introducing denoising convolutional networks prior to regular processing \cite{Chang21}, exploring different deep architectures \cite{Nie24}, or exploiting multiple optical paths or multiple wavelengths \cite{Ray23}.



\section{Hardware and methods}
\label{sec:hwsw}




The Pixel Watch 2 is the first Google device to exploit multiple optical paths combined with deep learning to combat motion artifacts (see Figure in abstract)\footnote{Fitbit Sense used multipath PPG, but with legacy FFT processing.}. 
We hypothesized that increasing the number of optical paths defined by the arrangement of green LEDs (525nm) and photodetectors (PDs) enables more effective separation of the pulse and motion signals than a single channel.
The TI AFE4500 analog front was configured to provide 10 PPG signals comprised of unique combinations of LEDs, PDs, and transimpedance amplifier gain settings.
This configuration was downselected experimentally from a wider initial set, taking into account power consumption and performance. 
The LED drive current is controlled by a closed-loop controller that aims 
to maximize
the amplifier gain while remaining in the dynamic range of the ADC despite fluctuations in ambient light and skin contact.
This forms a first strategy to minimize the effect of motion artifacts.
The signal is bandpass filtered between 0.5 and 12 Hz prior to the ADC. 
This range captures more of the pulse definition, which helps the model to better discern the HR signal.


After the ADC, heart rate is inferred from PPG and accelerometer input \textemdash used to capture the motion artifacts \textemdash by a deep network.
All sensor streams are downsampled to 25 Hz. 
The 10 optical channels are individually processed by a detrending and normalization block and a recursive least squares adaptive filtering and normalization block. 
These 20 optical and 3 motion inputs are fed to the deep network which comprises a 9-layer temporal convolutional feature extractor with filter count increasing from 40 to 56, designed to capture morphological features of the pulse wave.
This is followed by 6 temporally dilated convolutional blocks with exponentially increasing dilation rates ($d = 2^1, \dots, 2^6$) to integrate longer temporal context without the computational cost of recurrent or attention-based mechanisms.
The architecture is optimized to fit the system thermal and power envelope and has less than 300K parameters.
We addressed the problem as a classification problem. 
The output layer generates a probability distribution at 0.78 Hz over 94 classes corresponding to HR buckets of 2 BPM  width centered at 31, 33, ..., 217 BPM, and then interpolated to a 1 BPM resolution and 1 Hz output.
\replaced{The network provides an output even with high movement artifacts, but the probability distribution on the output layer can indicate decreased confidence.}{}
\section{Training and evaluation datasets}
\label{sec:dataset}



\subsection{Deep learning training dataset}
A key to model performance was training on large volume data.
We collated data from internal studies covering human activity recognition (573 participants), sleep studies (90 participants), acceptance testing (98 participants) and free-living data collection where users go about their everyday activities without constraints (593 participants), all with informed consent. 
Raw watch PPG data was recorded concurrently with a Polar H10 chest strap for ground truth.
We sampled this data to ensure a balanced representation of heart rate within pre-defined BPM brackets.
The resulting volume of data is 9319 hours from 962 participants split in 75\% for training (640 participants, 6864 hr) and 25\% for hyperparameter finetuning (322 participants, 2455 hr).

\subsection{Evaluation datasets}

Evaluation was performed in datasets that were not part of the development nor the finetuning of the PW2 algorithm.
The {\em Exercise} evaluation dataset is an in-house dataset during which 
participants
completed proctor-guided indoor and outdoor activities including remaining still, walking, running, rowing machine, exercise bike, high intensity interval training (HIIT), in order to understand how the device performs in different conditions (WIRB-Copernicus Group IRB Protocol No. 20233158, IRB Study No. 1357498, all with informed consent).
\replaced{This protocol has been consistently followed for multiple generations of devices and is inspired by CTA-2065.}{}
Not all activities were always performed, based on personal capabilities and preferences. 
%
%
Participants wore a PW2 randomly allocated to the dominant or non-dominant wrist 
and a Polar H10 ECG chest strap as the criterion device.
\replaced{Participants were instructed to wear the watch as per the product documentation \cite{GooglePixelWatchWear2026}.}{}
The data from both devices was acquired using a custom logging software at 1Hz.
The resulting analysis dataset comprises \replaced{229}{227} participants (136 men, 90 women, \replaced{1 non-binary and 2 not wanting to disclose}{and 1 non-binary/non-conforming}) and 248 hours of data. 
The participants' age ranged from 18 to 70 (mean=34.9, SD=10.1).
The skin color using the Monk scale ranged from 1-10 (mean=5.5, SD=2.1).
Wrist circumference
ranged from 130mm to 200mm (mean=161 mm, SD=13.5mm).
The breakdown of data across the different activities is reported in \cref{tab:dataset}.

\begin{table}[ht!]
\tiny
\centering
\caption{Exercise and Free-living dataset evaluation results (upper part of table) and independently published work evaluating Google devices (lower part of the table) and associated performance measures.}
\label{tab:dataset}
\begin{tabular}{|p{0.26\linewidth}|p{0.02\linewidth}|p{0.08\linewidth}||p{0.1\linewidth}|p{0.08\linewidth}|p{0.13\linewidth}|}
\hline
Exercise dataset & n & Duration [mn] & MAE (MAPE) & Bias [bpm] & LoA [bpm]\\
\hline
Still           &    227 &   1709 & 2.21 (2.20) & -1.29 &  -9.14 -   6.56\\
Walk            &    228 &   4304 & 1.83 (1.73) & -0.06 &  -7.54 -   7.43\\
Run             &    226 &   6338 & 2.56 (1.68) & -1.22 & -12.65 -  10.21\\
Indoor bike     &     97 &    933 & 1.04 (0.88) & -0.77 &  -6.21 -   4.67\\
Rowing machine  &    101 &    691 & 2.08 (1.62) & -1.34 &  -9.80 -   7.12\\
HIIT            &    115 &    909 & 2.39 (1.77) & -0.77 &  -9.88 -   8.35\\
\hline
Overall         &    229 &  14886 & 2.18 (1.71) & -0.84 & -10.34 -   8.66\\
\hline
\hline
Free-living  dataset &     27 &  63618 & 1.84 (2.53) & 0.45 &  -6.57 -   7.48 \\
\hline
\hline
\hline
\cite{Gorny17} ChargeHR low MVPA &  10 &  2509 & N/A & -4.24 & -17.4 - 8.95 \\
\cite{Gorny17} ChargeHR high MVPA &  10 &  2509 & N/A & -16.2 & -44.0 - 11.6 \\
\cite{Nissen22} Charge4 various activity & 23 & ~345 & 8.59 (9.74) & -1.66 & -26.75 - 23.43 \\
\cite{HajjBoutros23} Sense walking & 60 & N/A & N/A (4.4) & -1.26 & -13.44 - 10.91 \\
\cite{HajjBoutros23} Sense running & 60 & N/A & N/A (3.8) & -3.76  & -16.73 -9.22 \\
\hline
\end{tabular}
\end{table}





Another independently-collected dataset was used to corroborate results in free-living conditions with a slightly older and more diverse population. 
A total of 27 consenting participants (9 women and 18 men, aged from 49 to 73, mean=61.6, SD=7.0) from the Fenland Remote Assessment Study (ethics HBREC2025.12, University of Cambridge) wore a synchronized PW2 and an Actiheart 4 heart rate sensor 
for 48 hours whilst carrying out their normal daily activities. \replaced{Participants were free to tighten the watch strap according to their individual preferences.}{}
The Actiheart is a single-lead ECG-based device 
attached to the chest. 
This device employs a modified Pan-Tomkins peak detection algorithm which has been validated against multi-lead ECG \cite{Brage05}. 
The Actiheart data was post-processed using Gaussian process regression, yielding a single heart rate estimate every minute \cite{Stegle08}. This led to the removal of 1432 min of Actiheart data (1.9 \% of 76880 min of total data) when the 95\% prediction interval was larger than 5 BPM.
The PW2 data was averaged to match the same minutely resolution to give 63618 min or 82.7\% of the total recording duration.

\section{Results and discussion}
\label{sec:evaluation}

\begin{figure*}[hbt]
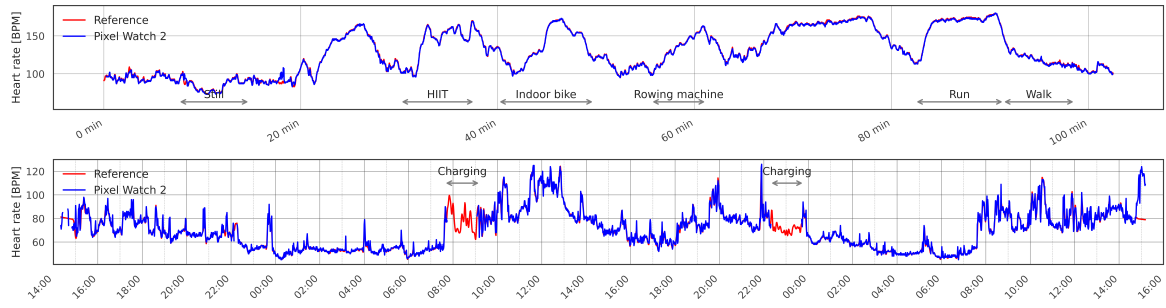

\centering
\includegraphics[width=.85\linewidth]{img/trace-act-compact-2910466-1200.png}\\
\includegraphics[width=.85\linewidth]{img/trace-fenland-4090448-arrow-1200.png}
\caption{Heart rate \replaced{recording}{cording} with the PW2 and reference device in the {\em Exercise} dataset (top) and the {\em Free-living} dataset (bottom).}
\label{fig:trace}
\end{figure*}

We show in~\cref{fig:trace} example data from the {\em Exercise} and the {\em Free-living} datasets. 
Activities of interest in the {\em Exercise} dataset are annotated, the non-annotated parts include other activities excluded from this evaluation. 
Note the rapid increase in heart rate at the onset of HIIT or Run, where the PW2 tracks the heart rate very well, showing the low latency of the deep learning algorithm compared to the criterion device.
In the free living dataset we can see two periods without PW2 data, which correspond to recharging the device.

\begin{figure*}[hbt]
\centering
\includegraphics[width=.85\linewidth]{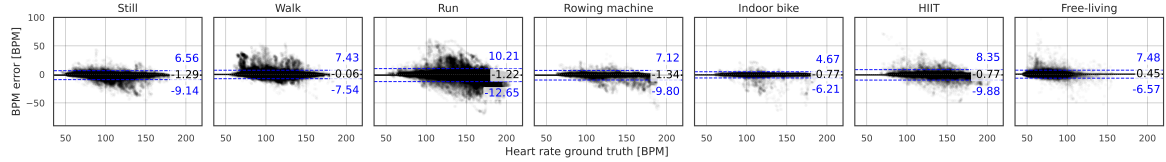}
\caption{Bland-Altman plots for the Exercise dataset (6 subplots on the left) and Free-living dataset (rightmost subplot).}
\label{fig:ba}
\end{figure*}

Let $HR_{w,i}$ be the $i$th HR measurement done with PW2
and $HR_{c,i}$ the measurement obtained concurrently with the criterion device. 
In the case of the {\em Exercise dataset}, $i$ refers to measurements at 1Hz, while for the {\em Free-living} dataset $i$ refers to measurements at 1 minute interval.
We compare $HR_{w,i}$ to $HR_{c,i}$ whenever data is available in both devices to derive bias, MAE, MAPE and 95\% LoA, as defined in \cite{Nissen22} which we report in table~\cref{tab:dataset} and~\cref{fig:ba}. 



Focusing on the LoA, we note that among the activities in the {\em Exercise} dataset the Indoor bike activity has the tightest LoA. 
This is due to fewer, smaller scale motion artifacts when holding the static bike handles. 
While one might have expected tighter error bounds during Still, this activity actually includes hand movements (e.g.,  pointing gestures instructed by the proctor, filling forms) and the user may either be seated or standing during this time. 

The {\em Free-living} dataset provides a more realistic depiction of performance and highlights that the model performs well in everyday life. 
The LoA in this dataset are somewhat tighter than the Overall LoA of the {\em Exercise} dataset and lie between Still and Walk in that dataset, possibly reflecting that a majority of the time participants engage in a combination of these activities.


In~\cref{tab:dataset} we list performance of prior Google products, including the first product with a HR sensor (ChargeHR) and the most recent trackers for which a validation study is available (Fitbit Charge4, a single optical path device, and Sense, a multipath device but using legacy FFT processing, both introduced in 2020).
The PW2 systematically achieves tighter LoA both in the {\em Exercise} and {\em Free-living} dataset than these prior products, even when comparing identical activities (e.g., Walk and Run with the Sense).





\section{Conclusion}
\label{sec:conclusion}

The PW2 is the first Google smartwatch with multipath optical hardware and deep learning-based heart-rate inference. 
We evaluated its performance on an {\em Exercise} dataset with \replaced{229}{227} participants and 248 hours of data and an independently collected {\em Free-living} dataset with 27 participants and 1000+ hours of data. 
This data volume is much larger than previous studies. 
The PW2 achieves 95\% Limits of Agreement (LoA) of -10.34 to 8.66 BPM in the {\em Exercise} and -6.57 to 7.48 BPM in the {\em Free-living} dataset.
These are much tighter LoAs than older Google products, especially in conditions of higher intensity physical activity (e.g., LoA of -26.75 to 23.43 BPM on the Charge 4 or LoA of -13.44 to 10.91 BPM in the Sense).

Deep learning \textemdash together with the high volume of training data (962 participants; 9,319 hours of data) \textemdash was the critical ingredient needed to fully leverage multipath PPG and significantly improve 
performance 
where previous classical DSP and Kalman filtering approaches fell short. 
Indeed, Charge 4 was a single path device with classical DSP, Sense was a multipath device with classical DSP, and the previous Pixel Watch 1 (PW1) was a single path device with deep learning, which also performed less well than PW2 in our internal testing.
Since the PW2, newer Pixel Watches have been introduced. They build incrementally on the PW2 architecture, notably with refinements to the electronics and optics, and further increase in the deep model size and training dataset volume.

\subsection{Lessons for wearable PPG HR sensor design}

There are several lessons for sensor design arising from this work. 
First, by using deep learning we were able to systematically explore optical path combinations  
to find the best performing ones with significantly reduced engineering effort: we did not design ad-hoc optical models, signal processing transforms, or machine learning features; instead we left this work to the convolutional feature extractors and only adjusted the model input dimension.

The benefit of the ML-based inference would be further magnified if one were to also optimize the physical placement of LEDs and PDs.
These combinations could be explored in an engineering smartwatch by having extra LEDs and PDs and electrically switching them over during long-duration data collection. 
The best placement and optical paths would be identified post-hoc by automatically training and evaluating models for each combination on the recorded dataset.

Training data volume is paramount. 
For this reason we conduct in-house data collection with the analog front-end (AFE) chip configured to acquire the broadest volume of multipath data possible. 

While our proposed model may seem large with 300K parameters, this is suitable to run on the Qualcomm Snapdragon W5 of the PW2 within the available power budget. Compute time is 40.11ms per inference, using 28mn of battery life out of 24 hours.

It is important to have very diverse users to assess effects of skin color, which affect PPG readings, or wrist-size, which affects attachment and motion artifacts. We encourage the community to consider this in future validation studies.


Finally, we learned that users often react more strongly to outliers than bias. 
\replaced{Internally we were also guided during algorithm development by a measure of ``perceived error'', which indicates the percentage of measurements deviating more than a threshold difference to the ground truth (e.g., 5BPM).}{In order to decide to release a product we therefore additionally developed a measure of ``perceived error'', representing how often users may perceive the measurement to be wrong, based on a configurable threshold difference to the ground truth.}

\section*{Acknowledgment}
\addcontentsline{toc}{section}{Acknowledgment}
\scriptsize
We thank all participants and fieldwork teams at Google and Cambridge. Funding: Google, UK Medical Research Council (MC UU 00006/4), Cambridge Biomedical Research Centre (203312).

\normalsize

\normalsize

%
%

%
%


\bibliographystyle{IEEEtran}
\bibliography{IEEEabrv,main}

\end{document}